\documentclass[12pt]{article}
\usepackage{amsmath}
\usepackage{slashed}
\usepackage{amsfonts}
\usepackage{amssymb}
\usepackage{tikz}
\usepackage{geometry}
\usepackage{bm}
\usepackage{setspace}
\usepackage{mathtools}
\usepackage{overpic}
\usepackage{graphicx}
\usepackage{cite}

\renewcommand\Im{\operatorname{Im}}
\renewcommand\Re{\operatorname{Re}}
\newcommand{\Tr}{\operatorname{Tr}}

\newcommand{\bmid}{\;\ifnum\currentgrouptype=16 \middle\fi|\;}

\newcommand{\colt}[4]{\mbox{\tiny $\begin{bmatrix}#1 & \vspace{0.05cm}\hspace{-0.2cm} #2\\#3 & \hspace{-0.2cm}  #4\end{bmatrix}$}}
\newcommand{\col}[2]{\mbox{\tiny $\begin{bmatrix}#1\\ \vspace{0.05cm} #2\end{bmatrix}$}}

\newcommand{\Sp}{\operatorname{Sp}}

\makeatletter
\renewcommand*\env@matrix[1][*\c@MaxMatrixCols c]{%
	\hskip -\arraycolsep
	\let\@ifnextchar\new@ifnextchar
	\array{#1}}
\makeatother

\DeclarePairedDelimiter\abs{\lvert}{\rvert}
\DeclarePairedDelimiter\norm{\lVert}{\rVert}

\makeatletter
\let\oldabs\abs
\def\abs{\@ifstar{\oldabs}{\oldabs*}}

\let\oldnorm\norm
\def\norm{\@ifstar{\oldnorm}{\oldnorm*}}
\makeatother

\numberwithin{equation}{section}

\usepackage[toc,page]{appendix}

\begin{document}

\begin{titlepage}
\begin{flushright}
	IPPP/17/6
	
	CERN-TH-2017-015
\end{flushright}
\samepage{
\setcounter{page}{0}
 \rightline{ }
\vfill
\begin{center}
\vspace{0.6in}
    {\Large \bf On exponential suppression of the cosmological constant in non-SUSY strings at two loops and beyond
    	\\} 
\vfill
\vspace{0.2in}
   {\large Steven Abel\footnote{E-mail address:
      {\tt s.a.abel@durham.ac.uk}}$^{\dagger,\ddagger}$ \, {\it and} ~
       Richard J. Stewart\footnote{E-mail address:
      {\tt richard.stewart@durham.ac.uk}}$^\dagger$\\}
   {$^\dagger$\it 
     IPPP and  Department of Mathematical Sciences, \\ Durham University, Durham, DH1 3LE, UK\\ }  
   {$^\ddagger$\it 
   	Theory Division, CERN, 1211 Geneva 23, Switzerland\\ }
\end{center}
\vfill
\begin{abstract}
  {\rm  
\noindent Two independent criteria are presented that together guarantee exponential suppression of the two-loop cosmological constant in non-supersymmetric heterotic strings. 
They are derived by performing calculations in both the full string theory and in its effective field theory, and come respectively from contributions that 
involve only physical untwisted states, and contributions that include orbifold twisted states. The criteria depend purely on the spectrum and charges, so a model that satisfies them will do so 
with no fine-tuning. An additional consistency condition (emerging from the so-called separating degeneration limit of the two-loop diagram) is that the one-loop cosmological constant must also be suppressed, by Bose-Fermi degeneracy in the massless spectrum. We comment on the effects of the residual exponentially suppressed one-loop dilaton tadpole, with the conclusion that the remaining instability would be under perturbative control in a generic phenomenological construction. We remark that theories of this kind, that have continued exponential suppression to higher orders, can form the basis for a string implementation of the ``naturalness without supersymmetry" idea.}
\end{abstract}
\vspace{2in}

\vfill
\smallskip}
\end{titlepage}

	\onehalfspacing
	\section{Introduction}
	
\begin{figure}[h]
	\centering
	\includegraphics[scale=0.88]{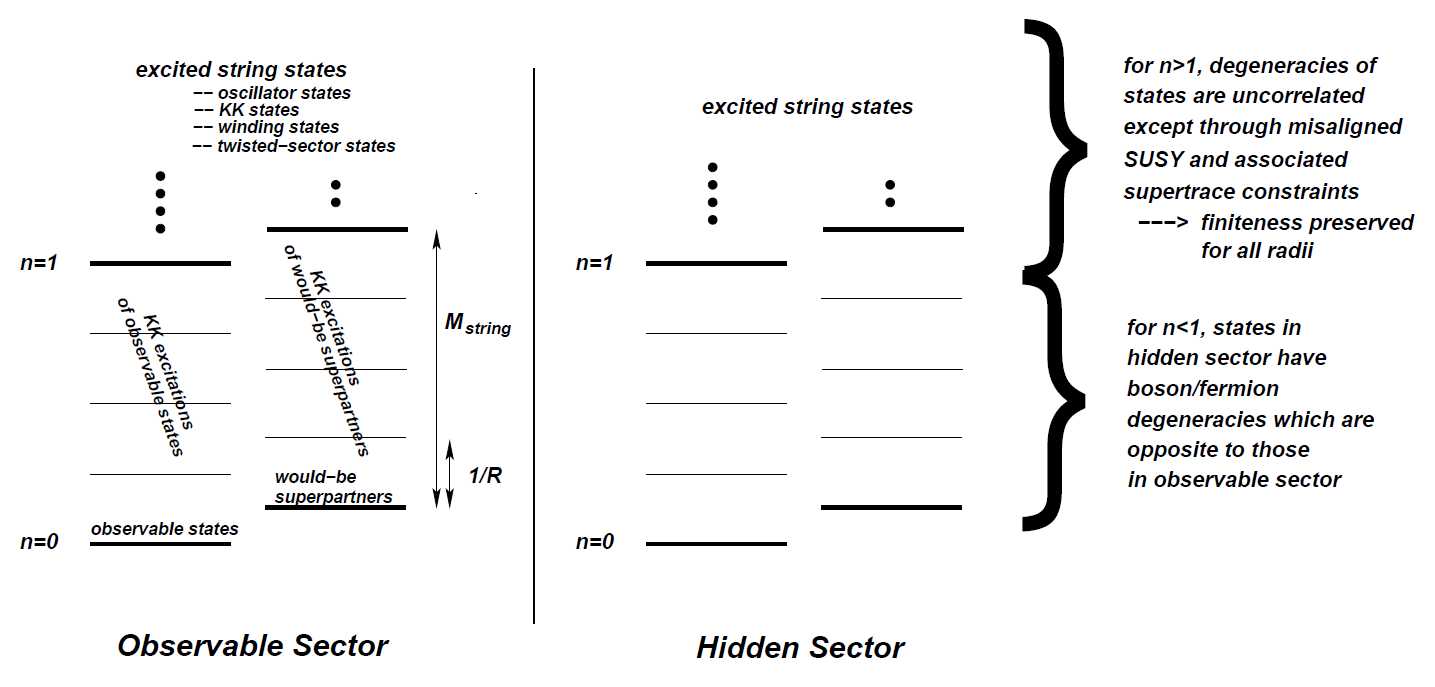}
	\caption{The spectrum of theories that satisfy Bose-Fermi degeneracy with a Standard Model-like light sector (reproduced from \cite{ADM}). As the Standard Model does not have Bose-Fermi degeneracy a cancelling hidden sector is inevitable, but note there is no supersymmetry in the spectrum. Models off this kind were constructed in \cite{ADM}.}
	\label{spectrum}
\end{figure}

	There has been interest recently in non-supersymmetric string theories, in which one might build the Standard Model (SM) directly. One particular object of focus has been the partial solution of the instability problems that generally arise in the absence of space-time supersymmetry (SUSY). In refs.\cite{ADM,KS,KKS} it was pointed out that a natural starting point for non-supersymmetric strings is a certain set of Scherk-Schwarz (SS) string models that have accidental Bose-Fermi degeneracy in their massless spectra. It is important to realise that this is a possibility even if the effective theory is entirely non-supersymmetric, and indeed such models were explicitly constructed in the string theory in \cite{ADM}. They have a 
	visible  spectrum that resembles that of the SM and a hidden sector whose Bose-Fermi non-degeneracy is equal and opposite to that of the SM, as shown in Figure \ref{spectrum}. In theories of this kind, 
	successive Kaluza-Klein (KK) levels are unable to contribute to the one-loop cosmological constant, which can only get 
	contributions from heavy winding modes, string excitation modes and also from non-level matched states. As these modes are all short-range, they are unable to explore the whole compact volume. Consequently, even if the compactification scale is only moderately large, their contribution to the cosmological constant (and hence destabilising dilaton tadpoles) is parametrically exponentially suppressed. 
	As an example, a supersymmetry breaking scale of say $1/R \sim 10^{14}$~GeV requires a string mass of $M_s\sim 10^{16}$ GeV to get the correct Planck scale (where $R$ represents the compactification scale and where the gravitino mass goes like $1/(2R)$). Even though the radius is then only $10^2$ 
	string lengths and the visible spectrum entirely non-supersymmetric, the cosmological constant is suppressed by an 
	astronomical factor,  $e^{-4\pi R M_s} \sim 10^{-546}$. (It is worth adding that more generally such a configuration 
	seems to be the only way to get an effective 4D {\it non-supersymmetric} theory. A {\it generic} 
	non-supersymmetric KK construction will be unstable and collapse on timescales of order $R$, thus it is never really four dimensional.
	Such a cosmological constant, generated entirely by heavy modes, also allows novel separations of finite UV and IR contributions to the potential \cite{SAA}.) 

	An open question is what happens at two-loops and beyond in such theories. Does the exponential suppression continue? Field theory intuition says that generic two-loop contributions will start to make their appearance, but it is conceivable that some kind of string ``miracle'' appears to save the day, or that a further subset of one-loop 
suppressed theories may have two-loop suppression in the cosmological constant as well. This paper shows by explicit calculation that (while we cannot rule out the former) the latter is highly likely. 
We derive two  criteria that define a sub-class of theories which continue to enjoy exponential suppression at two-loops. Like the one-loop case, this suppression is simply an accidental consequence of their particle content.

We should note that from a field theory point of view the possibility of such cancellation is surprising. At one-loop, some scalars and some fermions inevitably gain supersymmetry breaking masses of order the KK scale even though some others may remain exponentially light  \cite{ADM}: this then feeds back into the  cosmological constant which one generically expects to be of order the KK scale suppressed by a two-loop factor. However  the two-loop cosmological constant calculation in the full string theory {\it already contains within it} the one-loop shifts in the spectrum. Therefore another way to view the criteria we present, is that they precisely determine when the latter conspire to cancel in the two-loop cosmological constant. 

Our programme, and this entire approach, is reminiscent of the field theory ideas of refs.\cite{Jack:1989tv,Alsarhi:1991ji,Chaichian:1995ef,Masina:2013wja} which attempt to achieve {\em naturalness without supersymmetry}, by essentially extending the Veltman condition of ref.\cite{Veltman:1980mj} to all orders. 
Indeed, 
it is a remarkable fact that, thanks to the theorem of Kutasov and Seiberg \cite{Kutasov:1990sv}, 
non-supersymmetric string theories with $D=4$ whose cosmological constant vanishes at one-loop must also satisfy the ``field independent'' Veltman condition, namely ${\rm Str}(M^2)=0$ \cite{Dienes:1995pm,Dienes:1995nt}.
Hence although the object of study here is the cosmological constant, not the mass of some putative Higgs, there is a direct link.  However the string case goes even further than the field theory one: there are no freely adjustable couplings, since couplings are all either zero or one (or themselves exponentially suppressed by the volume), so 
there is  absolutely no fine-tuning involved. Theories either have the correct massless particle content or they do not. 

At the one-loop level, because of this connection to the Veltman condition, any model with vanishing cosmological constant can be thought of as a stringy UV completion of the scenario outlined in ref.\cite{Masina:2013wja}. Although we stress that the operator being considered here is the cosmological constant, the exact same procedure could be carried out for the Higgs mass-squared itself. This is discussed in more detail in ref.\cite{ADfuture}. 
In the models of ref.\cite{ADM}, this is achieved because a Scherk-Schwarz deformation preserves the Bose-Fermi degeneracy of the massless modes in all of their KK levels as well. 
In the logarithmically running low energy theory, one then assumes that the relevant scale at which such a relation should be applied is the compactification scale, above which the theory becomes extra dimensional. An important difference though is the motivation for imposing the condition at that scale which has nothing to do with SUSY being restored there, but rather the one-loop cosmological constant vanishing\footnote{Note that we cannot even say the theory becomes {\em approximately} supersymmetric at the scale $1/R$ because of the arguments presented in ref.\cite{ADM}: whilst at order $1/R$ the KK spectrum is indeed supersymmetric, the other stringy modes, in particular winding modes, manifestly break SUSY.}.

At the {\em two-loop} level, we will find as mentioned two rather different looking criteria for vanishing cosmological constant.  
The  criterion for the vanishing of the entirely untwisted contributions (that is diagrams whose propagators contain only the descendants of broken ${\cal N}=2$ supermultiplets) is a complicated combination of parameters 
(numbers of gauge bosons, gauginos, hypermultiplets and so forth) that essentially counts the two-loop effective field theory divergences. As we will demonstrate, this parameter is most easily extracted from the constant term in the ``$q$-expansion'' of the two-loop string partition function. By contrast diagrams that contain twisted loops (that is loops of twisted states that still appear in complete 
${\cal N}=1$ chiral supermultiplets) can vanish due to the cancellation of combinations of ``field dependent'' Veltman conditions. Such diagrams have a different dependence on the volume modulus from the entirely untwisted ones, so to avoid fine-tuning one has to impose a second independent criterion for the twisted states, of the form $\sum_{U} (-1)^{F_U}{\rm Tr}|Y_{UTT}|^2=0$ where $U$ stands for generic untwisted fields in the theory, and the trace is over the pairs of twisted states to which they couple, with tree-level coupling $Y_{UTT}$. This criterion is quite Veltman-like, but note that it is the {\em sum} over the Veltman conditions of all the twisted states that appears; we do not need to apply them individually. Furthermore the couplings are degenerate, so again the vanishing of this quantity is a question of particle content. 

An important aspect to bear in mind is that one requires an absence of gravitationally coupled products of one-loop divergences in order to produce the above criteria. This contribution 
would normally come from the so-called separating degeneration limit of the two-loop partition function, which we will discuss in some detail. Such terms are absent only if one has 
chosen a theory that already satisfies the criterion for the {\em one-loop} cosmological constant to vanish, namely massless Bose-Fermi degeneracy, $N_b^{(0)}-N_f^{(0)}=0$. Indeed, more generally one can see that at each order, a sensible criterion for continued suppression can only be 
achieved when the criteria for all the orders below are satisfied. 

The work contained in this paper naturally follows on from previous research into non-supersymmetric strings. The idea of Scherk-Schwarz SUSY breaking \cite{Scherk:1978ta} was first adapted to the string setting in refs.\cite{Kounnas:1989dk,Ferrara:1987es,Ferrara:1987qp,Ferrara:1988jx}, which introduced Coordinate Dependent Compactification (CDC). Subsequently, there has been extensive research into the one-loop cosmological constant \cite{Itoyama:1986ei,Itoyama:1987rc,Dienes:1994np,Dienes:1994jt,Dienes:1995gp,Dienes:1995pm,Dienes:1995nt,Dienes:2001se,Rohm:1983aq,Nair:1986zn,Ginsparg:1986wr,Moore:1987ue,Balog:1988dt,Dienes:1990qh,Dienes:1990ij,Kutasov:1990sv,KKS,KS,Harvey:1998rc,Kachru:1998yy,Blumenhagen:1998uf,Angelantonj:1999gm,Gaberdiel:1999jd,ShiuTye,IengoZhu,Faraggi:2009xy}, their finiteness \cite{Dienes:1994np,Dienes:1994jt,Dienes:1995gp,Dienes:1995pm,Dienes:1995nt,Angelantonj:2010ic}, how they relate to strong/weak coupling duality symmetries \cite{Bergman:1997rf,Bergman:1999km,Blumenhagen:1999ad,Blum:1997cs,Blum:1997gw,Faraggi:2007tj}, and ideas relating to the string landscape \cite{Dienes:2006ut,Dienes:2012dc}. The mechanism of CDC has been further developed in refs.\cite{Kiritsis:1997ca,Dudas:2000ff,Scrucca:2001ni,Borunda:2002ra,Angelantonj:2006ut} while phenomenological ideas have been explored further in refs.\cite{Nair:1986zn,Ginsparg:1986wr,Faraggi:2007tj,Lust:1986kj,Lerche:1986ae,Lerche:1986cx,Chamseddine:1988ck,Font:2002pq,Blaszczyk:2014qoa,Angelantonj:2014dia,Angelantonj:2015nfa,Blaszczyk:2015zta,Nibbelink:2016lzi}.  Additionally, solutions to the large volume ``decompactification problem" have been discussed in refs.\cite{Faraggi:2014eoa,Kounnas:2015yrc,Partouche:2016xqu,Kounnas:2016gmz}, while numerous other configurations of non-supersymmetric string models have been discussed in refs.\cite{Bachas:1995ik,Russo:1995ik,Tseytlin:1995zv,Nilles:1996kg,Shah:1997re,Sagnotti:1995ga,Sagnotti:1996qj,Angelantonj:1998gj,Blumenhagen:1999ns,Sugimoto:1999tx,Aldazabal:1999tw,Angelantonj:1999xc,Forger:1999ev,Moriyama:2001ge,Angelantonj:2003hr,Dudas:2004vi,GatoRivera:2007yi}, which have included the study of relations between scales in different schemes \cite{Antoniadis:1988jn,Antoniadis:1990ew,Antoniadis:1992fh,Antoniadis:1996hk,Benakli:1998pw,Bachas:1999es,Dudas:2000bn}.
 
The results we have found are a natural extension of this work, which leads one to speculate on the existence of three-loop and beyond cancellations, and whether there might be a universal condition for string theories that, like the one conjectured for field theory in ref.\cite{Jack:1989tv}, ensures cancellation to all orders. Conversely, it raises the possibility that imposing the requirement of continued exponential suppression to ever higher order could give interesting predictions for the particle content of the theory.  

	\section{Two-loop amplitudes}
	
	\subsection{The set-up in the $\vartheta$-function formalism}

Let us begin by collecting and digesting the necessary results for the calculation of the two-loop cosmological constant. Multiloop string calculations of the cosmological constant have been considered in the past in refs.\cite{VV2,ABRV,GIS,ISZ,Moore:1986sk,Atick:1987rk,Lechtenfeld:1987dm,Parkes:1988pq,Ortin:1991md,Lechtenfeld:1989ke,Lechtenfeld:1989ku,KS,KKS,IengoZhu}.
However, care is required from the outset as there are possible pitfalls.
In particular, one of the major difficulties in calculating string amplitudes beyond one-loop proved to be the integrating out of the  supermoduli.
If done incorrectly, computations of this type typically give ambiguous results that depend on the choice of gauge. For example, attempts were made in the past to determine the value of the two-loop vacuum amplitude for the non-supersymmetric models presented in refs.\cite{KS,KKS} (the so-called KKS models). The initial claim was that the cosmological constant is vanishing, but contradictory evidence was presented in ref.\cite{IengoZhu}. In fact both of these results suffered from the aforementioned issue of gauge dependence. A correct gauge-fixing procedure was later introduced in the work of refs.\cite{DP1,DP2,DP3,DP4}, and the computation was re-done in ref.\cite{ADP} with the conclusion that the two-loop contribution is indeed non-vanishing for the KKS models. It is these later papers that form the basis of our analysis. 
	
	For the type of non-supersymmetric model described in ref.\cite{ADM}, one does not actually expect the two-loop contribution to the cosmological constant to be identically zero. As described in the Introduction, the best one can achieve at one-loop is for it to be exponentially suppressed if the massless spectrum contains an equal number of bosons and fermions. Therefore we seek a similar suppression at higher loop order. 
	
	Note that as the main source of the cosmological constant (a.k.a. Casimir energy) in large volume Scherk-Schwarz compactifications is the massless spectrum, one might think it is preferable to 
	approach the entire problem from the perspective of the effective field theory. However at two loops, it is not always obvious how the string computation factorises onto the field theory diagrams. In addition one would have to perform an analysis in the effective softly broken supergravity, and there are certain purely string contributions, in particular the separating degeneration limit (of which more later), that one has to check. These issues are exacerbated by the fact that the string models typically have a large rank making it tedious to count states, and by the fact that one would in any case have to determine 
	all the tree-level couplings of the effective field theory. As we shall see, it is by contrast far easier to simply extract the coefficient of the relevant (constant) term from the $q$-expansion of the two-loop partition function. 
	
	The structure of two-loop superstring amplitudes is built upon the representation of the worldsheet by a super Riemann surface of genus two. Let us start with a brief outline of the essential properties of such surfaces, and as a warm-up exercise then perform the computation of the two-loop cosmological constant in an entirely supersymmetric theory. 
	
\begin{figure}[ht]
		\centering
		\includegraphics[scale=0.9]{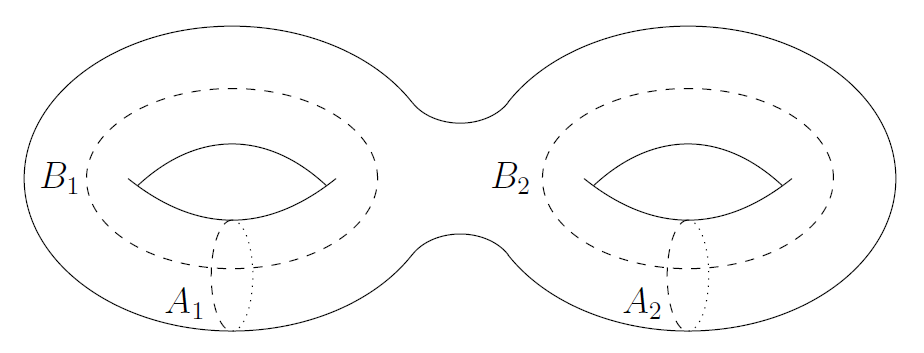}
		\caption{Canonical homology basis for genus 2.}
		\label{CanBas}
	\end{figure}

	Consider a super Riemann surface of genus $g$ with a canonical homology basis of 
	$A_I$ and $B_I$ cycles 
	as shown in Figure \ref{CanBas}. The period matrix $\Omega_{IJ}$ is given by holomorphic abelian 1-forms $w_I$ dual to the $A_I$-cycles such that
	\begin{equation}
	\oint_{A_I}\omega_J=\delta_{IJ}\,,\, \qquad\oint_{B_I}\omega_J=\Omega_{IJ}\,.
	\end{equation}
	In addition to the period matrix there is the {\it super} period matrix, $\hat{\Omega}_{IJ}$, which can be defined in a similar way, by integrating superholomorphic $1/2$ forms over the $A_I$ and $B_I$ cycles. 
	
	The supermoduli space $\mathfrak{M}_g$ of a genus $g$ super Riemann surface contains $3g-3$ even moduli and $2g-2$ odd moduli for $g\geq2$. 	
		Specialising to the case where $g=2$, the super period matrix gives a natural projection of the supermoduli space of a super Riemann surface onto the moduli space of a Riemann surface, and its 3 independent complex entries provide complex coordinates for the moduli space of even moduli, $\mathcal{M}_2$. The super period matrix can be expressed in a simple way in terms of the period matrix and, following the procedure of refs.\cite{DP1,DP2,DP3,DP4}, one can work in the so-called split gauge, which has the main advantage that 
	the period matrix and super period matrix are equivalent, and one can simply use  $\Omega_{IJ}$ to denote both. It can be parametrized by
	\begin{equation}
	\Omega=\begin{pmatrix}
	\tau_{11} & \tau_{12}
	\\
	\tau_{12} & \tau_{22}
	\end{pmatrix},
	\end{equation} 
	where $\tau_{11},\tau_{12}$ and $\tau_{22}$ are the complex variables corresponding to the three moduli (i.e. playing the same role as $\tau$ in the one-loop diagrams).
	To make the discussion widely accessible, we present the result (which derives from refs.\cite{DP1,DP2,DP3,DP4} after some work and carefully accounting for the measure) 
	in terms of  
	two-loop $\vartheta$-functions, the most natural extension of the standard one-loop formalism.

	For a genus 2 surface there are 16 independent spin structures, labelled by half-integer characteristics\footnote{Note that in our conventions, the spin structures are given as the transpose of those appearing in refs.\cite{DP1,DP2,DP3,DP4}}
	\begin{equation}
	\kappa=\begin{bmatrix} {\kappa'}\\ {\kappa''}\end{bmatrix} ,\qquad\kappa',\kappa''\in\left(0,\frac{1}{2}\right)^2\, ,
	\end{equation}
	where  $\kappa'$ is a 2-vector of spin structures on the $A_{I}$-cycles, and $\kappa''$ is a 2-vector of spin-structures on the $B_{I}$-cycles. 	

	The $\vartheta$-functions with characteristic $v$ are defined by 
	\begin{equation}
	\vartheta[\kappa](v,\Omega)\equiv\sum_{n\in\mathbb{Z}^2}\exp\{i\pi (n+\kappa')^t\,\Omega\,(n+\kappa')+2\pi i (n+\kappa')^t(v+\kappa'')\}\, .
	\end{equation}
	A given spin structure is said to be {\it even or odd} depending on whether $4\kappa'\cdot\kappa''$ is even or odd. For vanishing characteristics, $v=0$, all of the 6 {\it odd} spin-structure $\vartheta$-functions are identically zero (much like $\vartheta_{11}$ in the one-loop case), so that 
	\begin{equation}
\vartheta	\colt{\frac{1}{2}}{0}{\frac{1}{2}}{0}\,=\,\vartheta	\colt{0}{\frac{1}{2}}{0}{\frac{1}{2}}\,=\,\vartheta	\colt{\frac{1}{2}}{\frac{1}{2}}{\frac{1}{2}}{0}\,=\,\vartheta	\colt{\frac{1}{2}}{\frac{1}{2}}{0}{\frac{1}{2}}\,=\,\vartheta	\colt{\frac{1}{2}}{0}{\frac{1}{2}}{\frac{1}{2}}\,=\,  \vartheta\colt{0}{\frac{1}{2}}{\frac{1}{2}}{\frac{1}{2}}\,\stackrel{v\rightarrow0}{=}\, 0\,.
	\end{equation}
The even spin structures will be denoted generically with a $\delta$, and the even ones with a $\nu$: for example even $\vartheta$-functions will be written as $\vartheta[\delta]$. 

	After integrating over the supermoduli, enforcing the GSO projection and summing over spin structures, the cosmological constant for the {\it supersymmetric} heterotic string can be written  \cite{DP1,DP2,DP3,DP4} 
	\begin{equation}
	\label{eq:m1}
	\Lambda_{2-loop}=\int_{\mathcal{F}_2} \frac{d^3\Omega_{IJ}}{(\det\Im\Omega)^5}\frac{\Upsilon_8(\Omega)\overline{\Psi_8(\Omega)}}{\abs{16\pi^6\Psi_{10}(\Omega)}^2}\, ,
	\end{equation}
	where $d^3\Omega_{IJ}= d^2\tau_{11}d^2\tau_{12}d^2\tau_{22}$, and the integration is over the fundamental domain of the moduli, $\mathcal{F}_2$,  typically taken to be \cite{DGPR,PR,FP} \vspace{0.3cm}
	\begin{enumerate}
		\item \vspace{0.1cm}\hspace{1cm}$\displaystyle-\frac{1}{2}<\Re(\Omega_{11}),\Re(\Omega_{12})\Re(\Omega_{22})\leq\frac{1}{2}$\, ,
		\item \vspace{0.3cm}\hspace{1cm}$\displaystyle0<2\Im(\Omega_{12})\leq\Im(\Omega_{11})\leq\Im(\Omega_{22})$\, , 
		\item \hspace{1cm}$\displaystyle\abs{\det(C\Omega+D)}\geq1\ \ \forall\ \begin{pmatrix}
		A & B
		\\
		C & D
		\end{pmatrix}\in Sp(4,\mathbb{Z})$\, .
	\end{enumerate}
	
	The modular forms appearing in eq.(\ref{eq:m1}) are defined as follows. First  it is useful to define 
		\begin{equation}
	\Xi_6[\delta](\Omega)\equiv\sum_{1\leq i<j\leq 3}\langle\nu_i|\nu_j\rangle\prod_{k=4,5,6}\vartheta[\nu_i+\nu_j+\nu_k]^4(0,\Omega)\, .
	\end{equation}
	This expression uses the fact that any even spin structure can be written as the sum of  three odd spin structures, $\delta=\nu_{1}+\nu_{2}+\nu_{3}$; in the sum, $\nu_{4,5,6}$ are the remaining three odd spin structures, and
	\begin{equation}
	\langle\kappa|\rho\rangle\equiv\exp\{4\pi i(\kappa'\cdot\rho''-\rho'\cdot\kappa'')\}\, .
	\end{equation}
	  In term of $\Xi_6$ we then have    
	  \begin{align}
	  \Upsilon_8(\Omega) &=\sum_{\delta\text{ even}}\vartheta[\delta]^4(\Omega)\Xi_6[\delta](\Omega)\, , \nonumber \\
         {\Psi}_{10}({\Omega}) &=\prod_{\delta\text{ even}}{\vartheta}[\delta]^{2}(0,{\Omega})\, ,
	 \end{align} 
         where the product is obviously over even spin structures only.
        In the end the two-loop cosmological constant in a SUSY theory is of course zero,  as it should be; this is due to the genus two version of the abstruse identity, namely $\Upsilon_8=0$.

	\subsection{The Scherk-Schwarzed cosmological constant}
	
	Adapting the technology of the previous section, one can now start to build up the two-loop cosmological constant for the {\it non}-supersymmetric theories of ref.\cite{ADM}. These theories are constructed by taking a 6D theory in the free fermionic formulation and compactifying down to 4D on a 
	$\mathbb{T}_2 / \mathbb{Z}_2$ orbifold, breaking spacetime supersymmetry through a coordinate dependence in the compactification (CDC). This is the equivalent of the Scherk-Schwarz mechanism in string theory. 
	Sectors that are twisted under the final orbifolding remain supersymmetric under the deformation, and so their spectrum is unchanged. (Whenever we refer to ``twisted" or ``untwisted" this will always mean with respect to the final orbifolding.) At genus two there can be a twist associated with each loop, but the focus will mainly be on the totally untwisted sectors since twisted states are involved in a very restricted set of diagrams due to their remaining supersymmetric structure. 
	
It is worth elaborating on this last particular aspect before we start the calculation of the totally untwisted diagrams in earnest. One can proceed by constructing an extension of  the argument of refs.\cite{Ferrara:1987qp,ADM}. At one-loop the partition function of the ${\cal N}=0$ deformed theory  (whose orbifold action we shall denote by $g$) is decomposed as
\begin{align}
 {\cal Z}({\mathbf e}) &= \frac{1}{2} \left(   {\cal Z}^{0}_{0}({\mathbf e})-{\cal Z}^{0}_{0}({\bf 0})\right)\\
& + \frac{1}{2} \left(  {\cal Z}^{0}_{0}({\bf 0}) + {\cal Z}^{g}_{0}+ {\cal Z}^{0}_{g}+ {\cal Z}^{g}_{g}\right)\, ,
 \end{align}
where the indices represent the orbifold action on the $A$ and $B$ cycle. The Scherk-Schwarz phases on the world-sheet degrees of freedom are denoted by a vector ${\mathbf e}$. The only dependence on them is in the first totally untwisted term. The second term is (up to the factor of $1/2$)  the 
 partition function of the non-orbifolded and non-deformed ${\cal {N}}=2$ theory, while the second line is the partition function of an entirely undeformed ${\cal {N}}=1$ theory; both are zero, and hence only the first term can give a non-zero contribution to the cosmological constant. (So for example any ${\cal N}=2\stackrel{\mathbf e}{\rightarrow} {\cal N}=0$ un-orbifolded theory with Bose-Fermi degeneracy implies the existence of a chiral orbifolded ${\cal N}=1\stackrel{\mathbf e}{\rightarrow} {\cal N}=0$ theory that also has  
 Bose-Fermi degeneracy.)
 
Continuing to two loops, a similar decomposition would look like  
\begin{align}
4 {\cal Z}({\mathbf e}) &= {\cal Z}^{00}_{00}({\mathbf e})-{\cal Z}^{00}_{00}({\bf 0})+{\cal Z}^{0g}_{00}({\mathbf e})- {\cal Z}^{0g}_{00}({\bf 0})+\ldots   \nonumber \\
& \qquad+ \left( {\cal Z}^{00}_{00}({\bf 0}) + {\cal Z}^{0g}_{00}({\bf 0})+ {\cal Z}^{00}_{0g}({\bf 0})+ {\cal Z}^{g0}_{00}({\bf 0})+ {\cal Z}^{00}_{0g}({\bf 0})+\ldots \right. \nonumber \\
& \left. \qquad\qquad \qquad\qquad + {\cal Z}^{gg}_{00}+{\cal Z}^{g0}_{0g}+\ldots + {\cal Z}^{gg}_{gg}\, \right)  \, , 
 \end{align}
 where now of course there are two cycles. The bracket is the undeformed ${\cal N}=1$ theory and must vanish by supersymmetry, and the first term is the partition function for the entirely unorbifolded theory, representing contributions containing the untwisted fields only. Clearly the one loop argument would go through as before, were it not for the additional ${\mathbf e}$-dependent terms on the first line, which represent diagrams that have twisting on one pair of $A_I$, $B_I$ cycles, with the other pair of $A_I$, $B_I$ cycles remaining entirely untwisted. Such diagrams will be referred to as ``mixed'' diagrams. What remains is therefore to determine the contributions of the mixed diagrams at leading order, and the contribution from the entirely untwisted first term, ${\cal Z}^{00}_{00}({\mathbf e})$. It is these two different kinds of contribution that lead to the two criteria mentioned in the Introduction.  
 
 The former will be dealt with explicitly later, but for the moment let us now turn to the calculation for the entirely untwisted contribution which is (up to a factor) the cosmological constant of the un-orbifolded theory.  To define the sums over spin structures, the CDC and vector notation is the standard one, summarized in ref.\cite{ADM}. In particular dot-products  are the usual Lorentzian ones, while a separate sum over basis vectors ${\bf V}_a$ is understood; thus explicitly the collection of spin-structures in a particular sector are  $\alpha^I {\bf V}\equiv \alpha^I_a {\bf V}_a$ and  $\beta^I {\bf V}\equiv \beta^I_a {\bf V}_a$, with $a$ labelling the basis vectors and, recall,  $I=1,2$\, labelling the $A_I$ and $B_I$ cycles. The right- and left-moving fermions  have spin-structures denoted \[{\bf S}'_R= {\col{(\bm{\alpha V})'}{(\bm{\beta V})'}}_R\,\, ,\qquad  {\bf S}'_L=\col{\bm{(\alpha V)'}}{(\bm{\beta V})'}_L\,.\] The primes represent the shift due to the CDC deformation, that is 
	\begin{eqnarray}
	\left({\alpha}^I  {\bf V}\right)' &=& 	{\alpha^I} {\bf V} - n^I{\bm e}\nonumber \\ 
	\left(-{\beta}^I  {\bf V}\right)' &=& 	{-\beta^I}  {\bf V} + \ell^I{\bm e}\, ,
	\end{eqnarray}
	where $n^I=n^{1I}+n^{2I}$, $\ell^I=\ell_1^I+\ell_2^I$ and $n^{iI}$ are the winding numbers and  $\ell_i^I$ are the {\it dual}-KK numbers in the Poisson resummed theory.
	In the present context, there are 16 transverse right-moving real fermions and 40 transverse left-moving real fermions on the heterotic string (so that ${\bf S}'_{R/L}$ are vectors containing 16 and 40 different spin structures respectively).


	After a little work, the techniques of  ref.\cite{DP1,DP2,DP3,DP4} yield the two-loop cosmological constant expressed purely in the $\vartheta$-function formalism:
	\begin{equation}
	\label{eq:master}
	\Lambda_{2-loop}=\int_{\mathcal{F}_2} \frac{d^3\Omega_{IJ}}{(\det\Im\Omega)^3}\sum_{\{{\bm {\alpha}}^a,{\bm{\beta}}^a\}}\frac{\Gamma_{2,2}^{(2)}}{\abs{\Psi_{10}}^2}\,\tilde{C}' \col{\bm{\alpha}}{\bm{\beta}} \, \,\Xi_6\colt{\alpha^1s}{\alpha^2s}{\beta^1s}{\beta^2s}\,\,\prod_{i=1}^{16}\vartheta [S'_{R\, i}]^{1/2}\prod_{j=1}^{40}\bar{\vartheta}[{S'_{L\, j}}]^{1/2}\, ,
	\end{equation}
	where  $d^3\Omega_{IJ}= d^2\tau_{11}d^2\tau_{12}d^2\tau_{22}$ and where `$s$' denotes the non-compact space-time entries of the spin-structure vectors. 
	
	Let us describe the factors in detail. 
In addition to the self-evident fermion factors, 
	the compactification from 6D to 4D has introduced an extra factor of the two-loop Narain partition function for the two compact bosonic degrees of freedom, $\Gamma^{(2)}_{2,2}$. In its original {\it non}-Scherk-Schwarzed and {\it un}-Poisson resummed format it would look like 
	\begin{equation}
	\Gamma_{2,2}^{(2)}(\Omega;G,B)=\det\Im\Omega\sum_{(m_i^I,n^{iI})}e^{-\pi\mathcal{L}^{IJ}\Im(\Omega_{IJ})+2\pi im_i^In^{iJ}\Re(\Omega_{IJ})}\, ,
	\end{equation}
	where
	\begin{equation}
	\mathcal{L}^{IJ}=(m_i^I+B_{ik}n^{Ik})G^{ij}(m_j^J+B_{jl}n^{Jl})+n^{iI}G_{ij}n^{jJ},
	\end{equation}
	and where $G_{ij}$ and $B_{ij}$ are the usual metric and antisymmetric tensor respectively.
	 After introducing the CDC shift and performing a Poisson resummation on all of the $m$'s, it takes the form
	\begin{align}
	\begin{split}
	\label{eq:gam22}
	\Gamma_{2,2}^{(2)}=	
	T_2^2\sum_{\ell_i^I,n_i^I}\exp & \left\{-\frac{\pi T_2}{U_2\det\Im\Omega}
	\left[  |M_1^1+M_2^1U|^2 \Im{\tau_{22}} +  |M_1^2+M_2^2U|^2 \Im{\tau_{11}}    \right. \right. \\
	 - &\left.\left.\left.  \left( (M_1^1+M_2^1U) ({M}_1^2+{M}_2^2 {U})^* +c.c. \right) \Im{\tau_{12}}
 	\right]\right.\right. \bigg\} \times  e^{-2\pi i T (n_1^1 \ell_2^1 + n_1^2 \ell^2_2 - n_2^1 \ell_1^1 - n_2^2 \ell^2_1)}
	\end{split}
	\end{align}
	where 
	\begin{align}
	\begin{split}
	M_1^1&=\ell_1^1-n_1^1\tau_{11}-n_1^2\tau_{12}\, ,
	\\
	M_1^2&=\ell_1^2-n_1^2\tau_{22}-n_1^1\tau_{12}\, ,
	\\
	M_2^1&=\ell_2^1-n_2^1\tau_{11}-n_2^2\tau_{12}\, ,
	\\
	M_2^2&=\ell_2^2-n_2^2\tau_{22}-n_2^1\tau_{12}\, .
	\end{split}
	\end{align}
		We should point out that in the above equations and in what follows, we have lowered the '$i$' index on the winding numbers purely to simplify notation; they have {\it not} been lowered through the use of the metric $G_{ij}$. A word of warning is also required concerning the definition of the ${\{{\bm {\alpha}}^a,{\bm{\beta}}^a\}}$ summation in eq.(\ref{eq:master}): the partition function $\Gamma_{2,2}^{(2)}$ is of course a function of $\ell_i^I,n_i^I$, but now so are the ${{\bf S}'_L}$ and ${{\bf S}'_R}$  due to the CDC induced shift. Therefore one cannot {\it really} factor the summations as we appear to do above: everything to the right of $\Gamma_{2,2}^{(2)}$  is to be correctly included in the sum over $\ell_i^I,n_i^I$. However the case of ultimate interest is when the radii are moderately large, since as described in the Introduction we wish to determine the presence or otherwise of unsuppressed SS contributions to the vacuum energy. These can only correspond to $n^I=0\mod(2)$ as is evident from eq.(\ref{eq:gam22}), while we require at least one of the $\ell^{I=1,2}$ to be equal to $1\mod(2)$ to avoid cancellation by supersymmetry. The Poisson resummation could have been done for different choices of the $\ell^I$ separately but it would amount to the same result. The result is leading terms that carry the usual volume dependence but are otherwise not suppressed. Conversely the sub-leading terms coming from the non-zero $n^I$ modes would involve a simple generalisation of the saddle-point approximation used for the one-loop case in ref.\cite{ADM} leading inevitably to exponential suppression.

	The final ingredients in eq.(\ref{eq:master}) are the GSO projection phases, $\tilde{C} \col{\bm{\alpha}}{\bm{\beta}}$. These can be deduced from the 
fact that two-loop partition functions factorize onto products of two one-loop partition functions in a certain limit of moduli space, at which point the GSO coefficients must factorize as well \cite{ShiuTye,KLST}. Since the GSO coefficients are completely moduli independent, this factorization must hold everywhere. They can therefore be written as a product of the known genus one coefficients
	\begin{equation}
	\tilde{C}\col{\bm{\alpha}}{\bm{\beta}}=\tilde{C}\col{\alpha^1}{\beta^1}\tilde{C}\col{\alpha^2}{\beta^2}\, .
	\end{equation}
	As described in ref.\cite{AAM}, most generally these are functions of the structure constants $k_{ab}$, $k_{eb}$, $k_{ae}$ and $k_{ee}$, that take the following form  
	\begin{equation}
	\tilde{C}\col{\alpha^I}{\beta^I}=\exp\left[2\pi i\left(\ell^I k_{ee}n^I-\ell^I k_{eb}\alpha^I_b-\beta^I_a k_{ae} n^I      \right)\right]\exp\left[2\pi i(\alpha_a^I s_a+\beta_a^I s_a+\beta^I_a k_{ab}\alpha^I_b)\right]\, ,
	\end{equation}
	with the vector ${\bm e}$ assuming a projective role, completely analogous to that of the other basis vectors. 
	For the canonical assignment of structure constants for the CDC vector ${\bm e}$, there is no sector dependence in the phases, that is  
		\begin{equation}
	\tilde{C}\col{\alpha^I}{\beta^I}=\exp\left[2\pi i\left(  {\scriptstyle\frac{1}{2}}\,\ell^I\bm{e}^2 n^I-  \beta^I\bm{V} \cdot \bm{e} \,n^I \right)\right]\exp\left[2\pi i(\alpha_a^I s_a+\beta^I_as_a+\beta_a^Ik_{ab}\alpha_b^I)\right]\, .
	\end{equation}
	However, note that in eq.(\ref{eq:master}) we actually have $\tilde{C}' \col{\bm{\alpha}}{\bm{\beta}}$ rather than $\tilde{C} \col{\bm{\alpha}}{\bm{\beta}}$. This primed definition does not include the factors of $\exp[2\pi i(\alpha^I_as_a+\beta^I_as_a)]$ appearing in the above equations, which are effectively contained within $\Xi_6$ instead. 

	Eq.(\ref{eq:master}) is the ``master equation'' that provides our first criterion. It is straightforward to  check that it  has the correct modular properties under $Sp(4,\mathbb{Z})$ by considering the transformations given in eq.\eqref{ModT}. As we are about to see, one can also use it to determine the leading contribution to the cosmological constant by 
deduce the $q$-expansions, by inserting the explicit expressions for the two loop $\vartheta$-functions, in Appendix \ref{app:a}. Writing the cosmological constant as 
		\begin{equation}
	\label{eq:master3}
	\Lambda_{2-loop}=\int_{\mathcal{F}_2} \frac{d^3\Omega_{IJ}}{(\det\Im\Omega)^3}\aleph \, ,	\end{equation}
the criterion for vanishing untwisted contribution to the two-loop cosmological constant 
is then that the constant term in the $q$-expansion of 
	\begin{equation}
	\label{eq:master2}
	\aleph =\sum_{\{{\bm {\alpha}}^a,{\bm{\beta}}^a\}}\frac{\Gamma_{2,2}^{(2)}}{\abs{\Psi_{10}}^2}\,\tilde{C}' \col{\bm{\alpha}}{\bm{\beta}} \, \,\Xi_6\colt{\alpha^1s}{\alpha^2s}{\beta^1s}{\beta^2s}\,\,\prod_{i=1}^{16}\vartheta [S'_{R\, i}]^{1/2}\prod_{j=1}^{40}\bar{\vartheta}[{S'_{L\, j}}]^{1/2}\, ,
	\end{equation}
	vanishes.
	Note that $\aleph$ is a product of the measure and the partition function.
	
	\subsection{The $q$-expansion of $\aleph$}
\begin{figure}[h]
		\centering
		\includegraphics[scale=0.85]{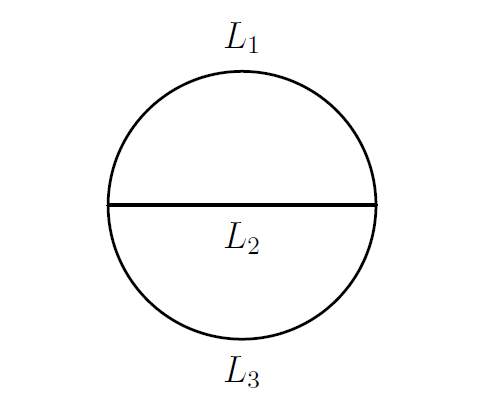}
		\caption{Generic sunset diagram for the two-point function.}
		\label{fig:sunset}
	\end{figure}

	Let us proceed to examine the $q$-expansions for the cosmological constant in certain limits, in particular the large radius limit.
       The general form of the integrand in the two-loop cosmological constant is 
	\begin{equation}
\aleph = \Gamma_{2,2}^{(2)}\sum_{\mathbf{a},\mathbf{b}\in\mathbb{Z}^3}C_{\mathbf{a}\mathbf{b}}q_1^{a_1}q_2^{a_2}q_3^{a_3}\bar{q}_1^{b_1}\bar{q}_2^{b_2}\bar{q}_3^{b_3}\,,
	\end{equation}
	where $a_i\geq-1/2$ and $b_i\geq-1$. 	
	It is useful to define $Y_{I=1..3}$ such that $\tau_{11}\equiv Y_1+Y_2,\ \tau_{12}\equiv Y_2,\ \tau_{22}\equiv Y_2+Y_3$ with $q_I=\exp\{2\pi i Y_I\}$. Letting $L_I=\Im(Y_I)$ so that
	\begin{equation}
	\Im\Omega=\begin{pmatrix}
	L_1+L_2 & L_2
	\\
	L_2 & L_2+L_3
	\end{pmatrix},
	\end{equation}
	the variables $L_1,L_2,L_3$ can be interpreted as Schwinger time parameters for the three propagators of the two-loop sunset Feynman diagram shown in Figure \ref{fig:sunset}. With this parametrization, $\det\Im(\Omega)=L_1L_2+L_2L_3+L_1L_3$, and the fundamental domain ${\cal F}_2$ restricts the variables so that $0<L_2\leq L_1\leq L_3$.

	By parameterizing the period matrix in this way, the $q_I$-expansion of $\aleph$ is symmetric with respect to the three $q_I$. It can be relatively straightforwardly evaluated.
	The $q$-expansion of $\Psi_{10}^{-1}$ is given by\footnote{Note that terms with a positive power of any of the $q_I$ are included in $\mathcal{O}(q_I)$. These terms can never contribute to the constant term in the full q-expansion.}
	\begin{equation}
	\frac{2^{12}}{\Psi_{10}}=\frac{1}{q_1q_2q_3}+2\sum_{I<J}\frac{1}{q_Iq_J}+24\sum_I\frac{1}{q_I}+\mathcal{O}(q_I)\, .
	\end{equation}
	The rest of $\aleph$ is model dependent and can be determined using the  $q_I$-expansions of the $\vartheta$-functions
	in Appendix \ref{app:a}. 
	
	As an example of the whole procedure we will consider an $SO(10)$ model that has massless Bose-Fermi degeneracy, and hence exponentially suppressed 
	cosmological constant at one-loop. The model is presented in Appendix \ref{so10}, where it is shown explicitly that in the SUSY theory (i.e. the theory {\it without} any CDC deformation) the two-loop cosmological constant vanishes. It is also shown there that the one-loop cosmological constant in the broken theory is exponentially suppressed because there is Bose-Fermi degeneracy at the massless level, and hence the constant term in the {\em one}-loop partition function is absent. 
	
	Recall that non-vanishing two-loop contribution to the cosmological constant comes from sectors in which at least one of $\ell^1$ and $\ell^2$ is equal to $1\mod{(2)}$. For example, if $\ell^1 = \ell^2 = 1 $, 	the $q$-expansion of $\aleph$ in the full non-SUSY $SO(10)$ theory is found to be 
	\begin{align}
	\label{dodo}
	\begin{split}
	\aleph&\propto\frac{1}{\abs{\Psi_{10}}^2}(q_1q_2q_3+\ldots)\left(1+\frac{1}{2}\bar{q}_1\bar{q}_2-\frac{33}{2}\bar{q}_1\bar{q}_3+\frac{1}{2}\bar{q}_2\bar{q}_3-116\bar{q}_1\bar{q}_2\bar{q}_3+\ldots\right)
	\\
	&=\frac{1}{\bar{q}_1\bar{q}_2\bar{q}_3}+\frac{2}{\bar{q}_1\bar{q}_2}+\frac{2}{\bar{q}_1\bar{q}_3}+\frac{2}{\bar{q}_2\bar{q}_3}+\frac{49}{2\bar{q}_1}+\frac{15}{2\bar{q}_2}+\frac{49}{2\bar{q}_3}-147+\mathcal{O}(q_I\bar{q}_J)\, .
	\end{split}
       \end{align}
       The terms with $\ell^1 = 1$ and $\ell^2 = 0 $, and with $\ell^1 = 0$ and $\ell^2 = 1 $ have the coefficients of $1/{{\bar q}_i}$ permuted but are otherwise identical. In particular the constant 
       term is the same. In total then, we find a non-vanishing constant piece, and conclude that this particular model gets a generic (i.e. not exponentially suppressed) contribution to the cosmological constant starting at two-loops. 
      
\subsection{Field theory factorization: identifying leading contributions}

\begin{figure}[t]
	\centering
	\includegraphics[scale=0.85]{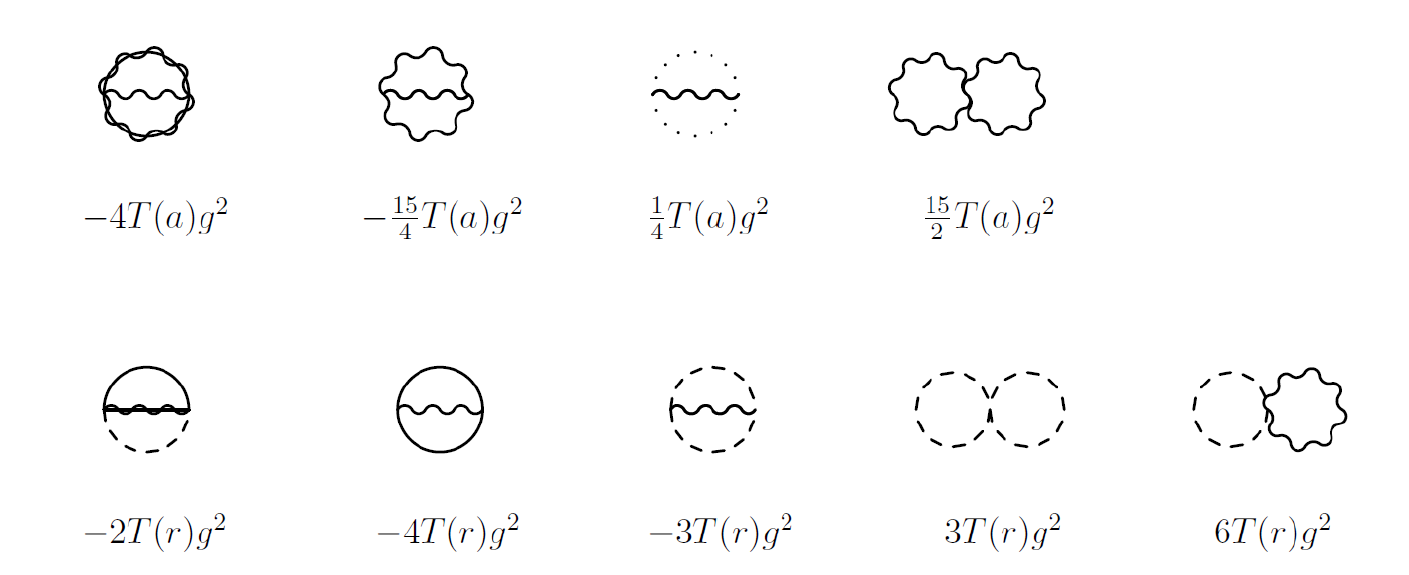}
	\caption{The Feynman diagrams for the two-loop cosmological constant in the effective ${\cal N}=2$ field theory of the untwisted sector with dashed lines indicating scalar components of hypermultiplets, solid lines fermionic components. Likewise ``photon'' lines represent the bosonic component of the gauge supermultiplet (i.e. vector plus scalar adjoint), while the gaugino lines represent the ${\cal N}=2$ gauginos. Leading order corrections (i.e. not exponentially suppressed) contributions are proportional to the sum over all these coefficients in the entire theory. In a supersymmetric theory the contributions vanish line by line as they should. In a Scherk-Schwarzed theory,  only those diagrams with {\it all masses unshifted} count (twice) towards the cosmological constant. Cancellation in a non-supersymmetric theory can achieved by choosing field content.}
	\label{fig:bubbles}
\end{figure}

Note that the constant piece in $\aleph$  includes various field theoretical contributions, not only the ones corresponding to the sunset topology. For reference the contributions in the field theory are displayed in figure \ref{fig:bubbles} in the parent ${\cal N}=2$ formalism. They can in principle be computed in the 6D field theory following ref.\cite{vonGersdorff:2005ce}. Given the complexity of the theories involved, and the fact that one would have to determine the spectrum {\em and} all the effective couplings, this would be an extremely arduous task, and it is actually much easier to simply determine the two-loop partition function directly as above. Nevertheless it is instructive  to see how the expression of eq.(\ref{dodo}) does indeed give the corresponding field theory contributions in the various degeneration limits. 

First note that for sufficiently large compactification volume the non-zero {\it winding} mode contributions are extremely exponentially suppressed compared to those with $n_i^I=0$. In addition the supersymmetric minimum for the CDC deformations is around $U_1=1$ as discussed in ref.\cite{SAA}. Expanding around this point and using eq.(\ref{eq:gam22}), the dominant contributions to the cosmological constant are  given by
	\begin{align}
	\label{urg}
	&\int_{\mathcal{F}_2}\frac{d^3\Omega_{IJ}}{(\det\Im\Omega)^3}\Gamma_{2,2}^{(2)}\bigg|_{n_i^I=0}\sum_{\mathbf{a},\mathbf{b}\in\mathbb{Z}^3}C_{\mathbf{a}\mathbf{b}}q_1^{a_1}q_2^{a_2}q_3^{a_3}\bar{q}_1^{b_1}\bar{q}_2^{b_2}\bar{q}_3^{b_3}\,\,\, \approx 
	\\
	&\, \int_{\sim 1}^\infty\int_{\sim 1}^{L_3}\int_{0}^{L_1}\frac{dL_2dL_1dL_3}{(\det\Im\Omega)^3}T_2^2\sum_{\ell_i^I,\,\mathbf{a}\in\mathbb{Z}^3}\exp\left\{-\,\frac{\pi T_2U_2}{\det\Im\Omega}\left[(\ell^1_2)^2L_3 +(\ell^2_2)^2L_1+(l_2^1-l_2^2)^2L_2 \right]\right. \nonumber 
	\\
	&\left. -\,\frac{\pi T_2}{U_2\det\Im\Omega}\left[ (\ell^1_1+\ell^1_2)^2L_3 +(\ell^2_1+\ell^2_2)^2L_1+(l_1^1+l_2^1-l_1^2-l_2^2)^2L_2 \right]\right\}C_{\mathbf{a}\mathbf{a}}
	\, e^{-4\pi(a_1L_1+a_2L_2+a_3L_3)}. \nonumber 
	\end{align}
	In the regions of the fundamental domain in which the real parts of the three moduli are integrated from $-1/2$ to $1/2$, the only non-zero contributions come from the physical states with $a_i=b_i\geq0$, and are given by the physical coefficients $C_{\bf aa}$. (This result is also a consequence of the fact that modular invariance requires $a_i-b_i\in\mathbb{Z}$.) The approximation sign is there because, as was also the case for one-loop integrals, there is a small region of the fundamental domain for which the integration over the real parts of the moduli does not extend over the full domain $-1/2<\Re(\Omega_{IJ})\leq1/2$. In this region, there is no level-matching and so unphysical states contribute to the vacuum amplitude. Nevertheless as in ref.\cite{ADM}, we find that the contributions from these unphysical states are also extremely exponentially suppressed compared to the both the massless contributions {\it and} the lowest lying string excitation mode contribution, provided that the compactification radii are sufficiently large.
	
	
	As per the previous subsection we are therefore interested in the value of $C_{\bf 00}$, the coefficient of the  constant piece giving leading order contributions. The important observation is that for these massless modes (with $a_1=a_2=a_3=0$)  the expression in eq.(\ref{urg})  has simply degenerated to the 4 dimensional field-theory result in the Schwinger formalism, so the  coefficient $C_{\bf 00}$ could also be calculated in the effective 6D$\rightarrow $4D Scherk-Schwarz field-theory.
 The relevant diagrams are shown together with the coefficients of their contribution to $C_{\bf 00}$ in figure \ref{fig:bubbles}, which are deduced from the calculations in ref.\cite{vonGersdorff:2005ce}. (Note that all coefficients are written for the fields as they decompose into boson or fermionic components of ${\cal N}=2$ multiplets.) 
	
	Different limits of the integral in eq.(\ref{urg}) generate {\it all} \,the field-theory diagrams in figure \ref{fig:bubbles}. In particular the ``double-bubble'' diagrams come from the region where $L_1, L_3\rightarrow  \infty$, while $L_2\gtrsim 1$. Explicitly in this limit, one still requires $a_1=a_2=a_3=0$ to avoid exponential suppression, but 
	 can everywhere replace $\det\Im\Omega \approx L_1L_3$. The $L_2$ integral then may be trivially performed (with its upper limit $L_1$ being effectively infinite). Taking for example $\ell_2^2=\ell_1^1=1$ in this limit results in an integral proportional to 
		\begin{align}
	\label{urg2}
	&\approx \,\,\,\,\,\, \int_{\sim 1}^\infty\int_{\sim 1}^{L_3}\frac{dL_1dL_3}{L_1^2L_3^2 }C_{\mathbf{0}\mathbf{0}}\exp\left\{-\,\frac{\pi T_2U_2}{L_3 }-\,\frac{\pi T_2}{U_2L_1}  \right\} \, ,\nonumber 
 	\end{align}
which (taking the upper limit $L_3\rightarrow \infty$ on the $L_1 $ integral) has the form of a product of two one-loop Poisson resummed Schwinger integrals in a KK theory with two extra dimensions. A more complete  
way to reach this conclusion would be to first go to the ``non-separating degeneration'' limit of ref.\cite{DP4}, i.e. $\tau_{22}\rightarrow i\infty$ with $\tau_{11},\tau_{12}$ fixed, and from there take $\tau_{11}\rightarrow i\infty$.

The field theory recipe for evaluating $C_{\bf 00}$ for the Scherk-Schwarzed string theories is therefore as follows: {\it retain in the list of two-loop diagrams only those that are {\it exactly} massless, meaning that the states on all  propagators  do not receive any CDC shift. Then $C_{\bf 00}$ is precisely {\it twice} the resulting sum of coefficients.}
 
The reasoning is straightforward and exactly mirrors what happens in the one-loop case. First recall that we are (for this calculation) considering only untwisted 
states in the diagrams of  figure \ref{fig:bubbles}. This implies that there is KK and ${\mathbf e}$ charge conservation at the vertices, which in turn implies that 
the CDC shifts {\it pairs} of either Fermion-Fermion or Boson-Boson masses on the sunset diagrams. The nett effect of such a shift is that the space-time statistics of an entire loop on the diagram is reversed, and consequently these diagrams contribute with an additional minus sign. Meanwhile the ``superpartner'' diagram (in which the space-time statistics {\it really is} reversed on that loop) is still present: hence a factor of two. 

In principle the sum of coefficients can vanish, and the important aspect that makes this possible is the coupling degeneracy, which is due to the underlying supersymmetry of the undeformed theory, and the ${\cal N}=2$ structure of the untwisted (i.e. SUSY breaking) sector. This is a well-known feature of effective string theories, but the crucial point here is that while at the level of the field theory a complete cancellation of contributions may seem like a miraculous tuning, at the level of the string theory it is merely a consequence of the particle content and the corresponding partition function  and measure  (and indeed there {\it are} no independent couplings). It is worth repeating that from this point of view (and in practice), it is far easier simply to work with the $q$-expansion of the string partition function, than to attempt to evaluate $C_{\bf 00}$ for the entire field theory.
 
\subsection{The separating degeneration limit}
\begin{figure}[h]
	\centering
	\includegraphics[scale=0.85]{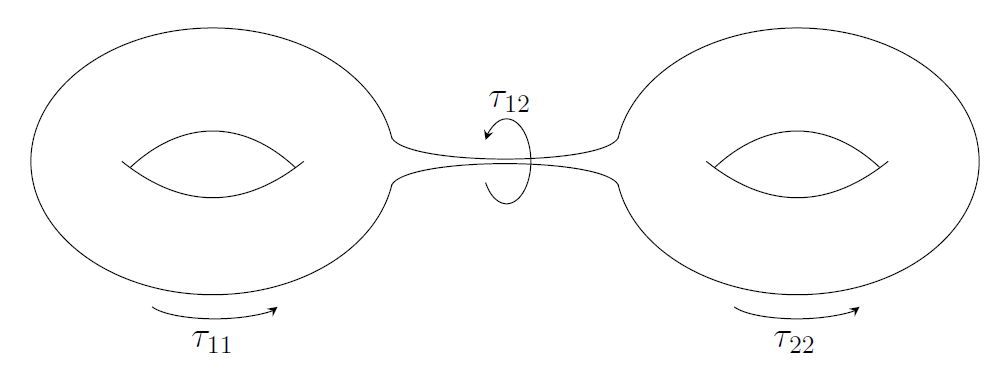}
	\caption{The separating degeneration limit.}
	\label{Septing}
\end{figure}

There is one limit that would not be covered by the field theoretic treatment described in the previous sub-section, namely the separating degeneration limit. For a two-loop string vacuum amplitude this corresponds to taking the limit $\tau_{12}\rightarrow0$ keeping $\tau_{11}$, $\tau_{22}$ fixed. This gives a Riemann surface that looks like two one-loop vacuum amplitudes connected by a long thin tube, as shown in Figure \ref{Septing}. The limits of various objects appearing in the two-loop cosmological constant are given by~\cite{DP4}
\begin{align}
\begin{split}
\vartheta[\mu_1,\mu_2](\Omega)&=\vartheta_1[\mu_1](0,\tau_{11})\vartheta_1[\mu_2](0,\tau_{22})+\mathcal{O}(\tau_{12}^2)\,,
\\
\vartheta[\nu_0,\nu_0](\Omega)&=-2\pi i\tau_{12}\eta(\tau_{11})^3\eta(\tau_{22})^3+\mathcal{O}(\tau_{12}^3)\,,
\\
\Xi_6[\mu_1,\mu_2](\Omega)&=-2^8\langle\mu_1|\nu_0\rangle\langle\mu_2|\nu_0\rangle\eta(\tau_{11})^{12}\eta(\tau_{22})^{12}+\mathcal{O}(\tau_{12}^2)\,,
\\
\Xi_6[\nu_0,\nu_0](\Omega)&=-3\cdot 2^8\eta(\tau_{11})^{12}\eta(\tau_{22})^{12}+\mathcal{O}(\tau_{12}^2)\,,
\\
\Psi_{10}(\Omega)&=-(2\pi\tau_{12})^22^{12}\eta(\tau_{11})^{24}\eta(\tau_{22})^{24}+\mathcal{O}(\tau_{12}^4)\,,
\end{split}
\end{align}
where $\mu_{1,2,3}$ and $\nu_0$ are the three even and unique odd genus 1 spin structures respectively, while the genus two Narain lattice $\Gamma_{2,2}^{(2)}$ splits into a product of two genus one Narain lattices. 
Therefore the full two-loop cosmological constant in the separating degeneration limit takes the form
\begin{align}
\begin{split}
\Lambda&=\int\frac{d^2\tau_{11}d^2\tau_{22}d^2\tau_{12}}{(\Im(\tau_{11})\Im(\tau_{22}))^{3}}\sum_{\{\alpha^i,\beta^i\}}\tilde{C}\col{\bm{\alpha}}{\bm{\beta}}\frac{1}{2^{18}\pi^4\abs{\tau_{12}}^4}\frac{1}{\eta(\tau_{11})^{12}\eta(\tau_{22})^{12}\bar{\eta}(\tau_{11})^{24}\bar{\eta}(\tau_{22})^{24}}
\\
&\times\Gamma_{2,2}^{(1)}(\tau_{11})\Gamma_{2,2}^{(1)}(\tau_{22})\prod_{\eta\in F_R'}\vartheta_{1}^{1/2}\col{(\alpha^1V)'}{(\beta^1V)'}\vartheta_{1}^{1/2}\col{(\alpha^2V)'}{(\beta^2V)'}\prod_{\tilde{\phi}\in F_L'}\bar{\vartheta}_{1}^{1/2}\col{(\alpha^1V)'}{(\beta^1V)'}\bar{\vartheta}_{1}^{1/2}\col{(\alpha^2V)'}{(\beta^2V)'}+\mathcal{O}\left(\frac{1}{\tau_{12}}\right),
\end{split}
\end{align}
which is essentially two one-loop vacuum amplitudes connected by a divergent propagator. We therefore make the 
crucial conclusion that the separating
degeneration limit contains the divergence due to any uncancelled one-loop dilaton tadpoles. In general, i.e. at higher loop order, 
one expects such terms to always be present. That is at $n$-loop order, any uncancelled tadpoles from the $(n-1)$-loop theory will 
contribute to divergences in the cosmological constant. Thus if the one-loop partition function
has Bose-Fermi degeneracy, these terms are a divergence multiplied by an exponentially suppressed coefficient. 

One may confirm that the same conclusion is arrived at using the full $q$-expansion in the separating degeneration limit. First of all in this limit  we have
\begin{equation}
\frac{2^{12}}{\Psi_{10}}=-\frac{1}{(2\pi\tau_{12})^2}\left(\frac{1}{q_1q_3}+\frac{24}{q_1}+\frac{24}{q_3}+576+\mathcal{O}(q_I)\right).
\end{equation}
Returning to the non-SUSY $SO(10)$ model with massless Bose-Fermi degeneracy given in Appendix \ref{so10}, for the untwisted sector with $\ell^{1,2}$ odd, the leading term in the $q$-expansion of the partition function after summing over spin structures is given by
\begin{align}
\begin{split}
\aleph&=\frac{1}{\abs{\Psi_{10}}^2}\left(-\frac{1}{4}+6q_1+6q_3-144q_1q_3+\ldots\right)\left(\bar{q}_1\bar{q}_2+\ldots\right)
\\
&=\frac{1}{\abs{2\pi\tau_{12}}^4}\left(-\frac{576}{4}+6\cdot 24+6\cdot 24-144+\mathcal{O}(q_I)\right)\left(1+\mathcal{O}(\bar{q}_I)\right)
\\
&=0+\frac{\mathcal{O}(q_I)\left(1+\mathcal{O}(\bar{q}_I)\right)}{\abs{2\pi\tau_{12}}^4}\, .
\end{split}
\end{align} 
The constant term has vanished as expected in this limit, for this model.

 \subsection{Comments on the effect of the one-loop tadpole}
 For the class of non-SUSY string models that we are considering in this paper, it is known that at one-loop order there is an exponentially suppressed but non-zero dilaton tadpole. If this tadpole is left uncancelled, then as we saw in the previous section, it can contribute through the separating degeneration as a divergence in the two-loop cosmological constant. It is well known that infrared divergences can appear in this degeneration \cite{Witten:2012bh,Witten:2013cia,Sen:2015cxs}, however, our experience from QFT is that these divergences typically arise because we are asking the wrong questions. As we have learned from QFT, what one should in principle do is stabilise the theory in the correct one-loop vacuum so that the tadpole is effectively cancelled. The two loop separating degeneration divergence  would  then be 
 seen to 
 be merely an artifact that disappears if we perform this procedure. It might also be the case that one could live with the tadpole and have a dynamical cosmologically evolving background as in ref. \cite{Dudas:2004nd}. These issues have also been discussed in refs. \cite{Angelantonj:2007ts,Kitazawa:2008hv,Sen:2015uoa}.
 
 In generic non-supersymmetric string models the dilaton tadpoles can be large. Any attempt to cancel the tadpole through a background redefinition would require such a large shift that it is highly unlikely that the new vacuum bears any resemblance to the original, thereby negating any positive phenomenological aspects of the originally constructed model. The key point about the specific types of models we consider here is that the dilaton tadpoles are exponentially suppressed. If one were to employ a background redefinition, the shift to achieve this should be sufficiently small so as not to result in any appreciable alteration in the phenomenological properties, including the spectrum of the massless states. If this were not the case then clearly there would be a problem, since the construction of models with suppressed cosmological constants is dependent on a careful cancellation of bosonic and fermionic massless degrees of freedom at one-loop order. In theory one is able to perform this background shift at the string theory level (see ref.\cite{Pius:2014gza}), however in practice this would be rather involved.
 
 An alternative argument is built around balancing the one-loop tadpole itself against another contribution as in ref. \cite{SAA} where the mechanism is
 incorporated in the effective supergravity theory, and of course should not itself result in a large cosmological constant. In a framework that is completely stable, where the dilaton tadpole is cancelled, the divergent contribution to the two-loop cosmological constant should then vanish, while crucially the remaining contributions remain unaltered. For the models which contain a bose-fermi degeneracy, the potential can be written as
 \begin{equation}
 V=V_{\text{IR}}+V_{\text{UV}},
 \end{equation}
 where $V_{\text{UV}}$ is computed in the full string theory while $V_{\text{IR}}$ arises from non-perturbative effects in the effective field theory. The key point is that because $V_{\text{UV}}$ comes from the contribution of heavy modes only, it is independent of the low-energy IR physics. Therefore, we can introduce some stabilising mechanism in the IR to cancel the UV contribution, and provided this does not alter the masses of states in any way that is not exponentially suppressed, then the massless spectrum will remain unchanged.
 
 A full treatment of the tadpole is beyond the scope of this paper and so we leave a complete study of the dynamics to future work. With this in mind, we assume it is fact consistent to study the cosmological constant in our naive vacuum, with the knowledge that the conditions on the structure of the massless spectrum that guarantee exponential suppression will still be satisfied after the shift to the correct vacuum. We emphasise that this would not be the case without exponential suppression of the one-loop tadpole. Those theories would undergo large shifts in the metric upon finding their true vacua, and any putative dilaton stabilisation would most likely be completely invalidated in the process, along with any two-loop discussion.

\subsection{Suppression of the ``mixed'' diagrams}

This completes the derivation and discussion of the first criterion for vanishing two-loop cosmological constant. It remains to consider the contributions with one untwisted propagator and 
two twisted ones, i.e. the mixed diagrams.  In the untwisted sector, the compactification from 6D to 4D resulted in the inclusion of the two-loop Narain partition function for the two compact bosonic degrees of freedom. This term meant that, for sufficiently large compactification radii, contributions to the cosmological constant from non-level matched states (including the proto-graviton) were exponentially suppressed compared to contributions from both massless states and the lowest lying string excitation modes. By contrast, for the twisted sectors, the partition function for the two compact bosonic degrees of freedom is given by \cite{ADP,DVV}
	\begin{equation}
	\mathcal{Z}[\epsilon]=\mathcal{Z}^{\text{qu}}[\epsilon]\sum_{(p_L,p_R)\in\Gamma}\exp\left\{\pi i\left(p_L^2\tau_\epsilon-p_R^2\bar{\tau}_\epsilon\right)\right\}
	\end{equation}
	where $\tau_\epsilon$ is the Prym period and 
	\begin{equation}
	\mathcal{Z}^{\text{qu}}[\epsilon]=\abs{\frac{\vartheta[\delta_i^+](0,\Omega)\vartheta[\delta_i^-](0,\Omega)}{Z(\Omega)^2\vartheta_i(0,\tau_\epsilon)^2}}
	\end{equation}
	where $Z(\Omega)$ is the partition function for two bosonic degrees of freedom in the uncompactified theory. 
	
	For twisted sectors involving some twist on only one of the two loops we anticipate that the cosmological constant may still receive a non-zero contribution. First we can see that again it is the massless states which provide the dominant contributions to the cosmological constant, while massive states receive exponential suppression after integrating over the real parts of the three moduli as before. The contributions from non-level matched (i.e. unphysical) states are also exponentially suppressed (for sufficiently large compactifiction radii), despite the fact these sectors do not include the two-loop Narain partition function. Instead, in these sectors there is the factor,
	\begin{equation}
	\Gamma_{2,2}^{(1)}(\tau_\epsilon)=\sum_{(p_L,p_R)}\exp\left\{\pi i\left(p_L^2\tau_\epsilon-p_R^2\bar{\tau}_\epsilon\right)\right\}
	\end{equation}
	which just has the form of a one-loop Narain partition function involving the Prym period $\tau_\epsilon$. As usual we can perform a Poisson resummation giving
	\begin{equation}
	\Gamma_{2,2}^{(1)}(\tau_\epsilon)=\frac{T_2}{\tau_\epsilon}\sum_{\vec{l},\vec{n}}\exp\left\{-\frac{\pi T_2}{\tau_\epsilon U_2}\abs{l_1-n_1\tau_\epsilon+(l_2-n_2\tau_\epsilon)U}^2\right\}.
	\end{equation}
	In order to show that the unphysical states are suppressed even in the twisted sectors, we make use of the fact that there is a relation between the  Prym period $\tau_\epsilon$ and the period matrix $\Omega$. The Schottky relations state that for any $i,j=2,3,4$
	\begin{equation}
	\frac{\vartheta_i(0,\tau_\epsilon)^4}{\vartheta_j(0,\tau_\epsilon)^4}=\frac{\vartheta[\delta_i^+](0,\Omega)^2\vartheta[\delta_i^-](0,\Omega)^2}{\vartheta[\delta_j^+](0,\Omega)^2\vartheta[\delta_j^-](0,\Omega)^2}.
	\end{equation}
	In the notation above, for any given twist $\epsilon\neq0$, there are 6 even spin structures $\delta$ where $\delta+\epsilon$ is also even. These 6 spin structures are denoted $\delta_i^+$ and $\delta_i^-$, for $i=2,3,4$, where $\delta_i^-=\delta_i^++\epsilon$. The region of moduli space where there is no level-matching is when $L_1,L_2,L_3$ are all sufficiently small and are at most $\mathcal{O}(1)$. When the imaginary parts of the three moduli are small, the Schottky relations tell us that $\Im(\tau_\epsilon)$ is also small (while it is large when both $L_1$ and $L_3$ are sufficiently large) and so by considering the Poisson resummed form of $\Gamma_{2,2}^{(1)}(\tau_\epsilon)$ we see that small values of $\tau_\epsilon$ result in exponential suppression.
	
		What remains therefore are the diagrams with a twisted loop and an untwisted propagator containing only physical states. (Due to the $\mathbb{Z}_2$ orbifold, there can only be either $UUU$ or $TTU$ vertices in the superpotential of the unbroken theory, and hence no diagrams with a single twisted propagator.) The coefficients of these diagrams can be easily evaluated in the field theory. 
		The integral for a loop of fermions of mass $m_1$ and $m_2$ coupling to a scalar are of the form
		\begin{equation}
		\Sigma(k^2)=-i\int\frac{d^4q}{(2\pi)^4}\frac{\slashed{q}+m_1}{q^2-m_1^2}\frac{(\slashed{q}+\slashed{k})+m_2}{(q+k)^2-m_2^2}\, .
		\end{equation}
		We can assume one mass to be zero, and first consider the fermion as the KK states. Thus we have to consider the Euclideanized integrals
		\begin{equation}
		I_{T_fT_sU_f}=2\int\frac{d^4q}{(2\pi)^4}\int\frac{d^4k}{(2\pi)^4}\frac{q\cdot(q+k)}{q^2k^2}\frac{1}{(q+k)^2+m_f^2}\, .
		\end{equation}
		We also have the case where the scalar is the KK state which involve the integral
		\begin{equation}
		I_{T_fT_fU_s}=\int\frac{d^4q}{(2\pi)^4}\int\frac{d^4k}{(2\pi)^4}\frac{q\cdot(q+k)}{q^2}\frac{1}{(k^2+m_s^2)}\frac{1}{(q+k)^2}.
		\end{equation}
		These diagrams will come with a coefficient $\Tr Y_{UTT}^2$ where $Y_{UTT}$ is the tree-level UTT Yukawa coupling in the superpotential; it takes the value $\sqrt{2}g_{YM}$ or $0$ depending on whether the charges are conserved at the vertex. The double-bubble diagrams (for Yukawas) will have the same coefficient  with a minus sign
		\begin{equation}
		J_{T_sT_s}=-\int\frac{d^4q}{(2\pi)^4}\int\frac{d^4k}{(2\pi)^4}\frac{1}{q^2}\frac{1}{k^2}.
		\end{equation}
		\begin{equation}		J_{T_sU_s}=-2\int\frac{d^4q}{(2\pi)^4}\int\frac{d^4k}{(2\pi)^4}\frac{1}{q^2}\frac{1}{k^2+m_s^2}.
		\end{equation}
		In the untwisted sector, it is possible to show that the sunset diagrams can be reduced to the form of scalar double-bubble diagrams by basic manipulation \cite{vonGersdorff:2005ce}. However, similar manipulations do not produce the same result in the twisted sectors and so we must evaluate the sunset diagrams as they are. Using the Schwinger formula
		\begin{equation}
		\frac{1}{A^\nu}=\frac{1}{\Gamma(\nu)}\int_0^\infty dyy^{\nu-1}\exp(-yA),\qquad\Re(A)>0,
		\end{equation}
		and the integrals
		\begin{equation}
		\int\frac{d^4q}{(2\pi)^4}q^{2n}\exp(-\alpha q^2)=\frac{\Gamma[2+n]}{\alpha^{2+n}16\pi^2}\, ,
		\end{equation}
		we find that the sunset diagrams can be written in the following form, where either $m_s=0$ if the single untwisted propagator is a fermion, or $m_f=0$ if it is a scalar:
		\begin{equation}
		I=-\frac{i}{(16\pi^2)^{2}}\int_0^\infty dy_1dy_2dy_3e^{-y_3m_s^2-y_2m_f^2}\times\frac{2y_3}{(y_1y_2+y_1y_3+y_2y_3)^{3}}.
		\end{equation}
		The above integral has UV divergences when at least two of the Schwinger parameters $y_1,y_2,y_3$ approach zero. Therefore, when we come to evaluate these diagrams later we will introduce a regulator $e^{-N\left(\frac{1}{y_2}+\frac{1}{y_3}\right)}$. 
		
		\subsubsection{Figure 8 diagrams}
		We may proceed to calculate the relevant integrals in a similar manner to refs.\cite{ArkaniHamed:2001mi,Ghilencea:2001bv}. In the untwisted sector the scalar figure 8 diagram is proportional to $J(m_{B_l}^2)^2$ where
		\begin{equation}
		J(m_{B_{m}}^2)=\sum_{m_i\in\mathbb{Z}}\int\frac{d^4p}{(2\pi)^4}\frac{1}{p^2+m_{B_m}^2}.
		\end{equation}
		We need to consider the case with two compact dimensions with radii $R_1$ and $R_2$. We will begin by considering the supersymmetric case in order to verify cancellation between all diagrams. For the scalar mass we therefore have 
		\begin{equation}
		m_{B_\ell}^2=\frac{4m_1^2}{R_1^2}+\frac{4m_2^2}{R_2^2}\, ,
		\end{equation}
		where $m_1$ and $m_2$ are Kaluza-Klein numbers.	Therefore, again using the Schwinger formula and integrating over the momentum $p$ we obtain
		\begin{equation}
		J(m_{B_{\ell}}^2)=\frac{1}{16\pi^2}\sum_{m_i\in\mathbb{Z}}\int_0^{\infty}dt\frac{1}{t^2}e^{-4\left(\frac{m_1^2}{R_1^2}+\frac{m_2^2}{R_2^2}\right)t}
		\end{equation}
		To proceed with the calculation we introduce a regulator $e^{-N/t}$, allowing us to interchange the order of summation and integration. From there we can perform a Poisson resummation on the KK numbers and finally obtain
		\begin{align}
		\begin{split}
		J(m_{B_{\ell}}^2)&=\frac{1}{16\pi^2}\int_{0}^\infty dt\frac{1}{t^2}\frac{\pi R_1R_2}{4t}\sum_{\ell_i\in\mathbb{Z}}e^{-\frac{\pi^2}{4t}\left(R_1^2\ell_1^2+R_2^2\ell_2^2\right)}e^{-\frac{N}{t}}
		\\
		&=\frac{1}{16\pi^2}\left[\frac{\pi R_1R_2}{4N^2}-\frac{4E\left(iU_2,2\right)}{\pi^3R_1R_2}+\frac{32NE\left(iU_2,3\right)}{\pi^5R_1^2R_2^2}\right]
		\end{split}
		\end{align}	
		where $U_2=R_2/R_1$ and $E(U,n)$ is the real analytic Eisenstein series with $U=U_1+iU_2$
		\begin{equation}
		E(U,n)=\sideset{}{'}\sum_{\ell_1,\ell_2}\frac{U_2^n}{\abs{\ell_1+\ell_2U}^{2n}}.
		\end{equation}
		For a twisted loop there are no associated KK states and so we only have the contribution from the massless state. In this case we simply have $J=\frac{1}{16\pi^2N}$ and so for the figure 8 diagram with a single twisted loop we find
		\begin{equation}
		J_{T_sU_s}=\frac{1}{(16\pi^2)^2}\left[\frac{\pi R_1R_2}{4N^3}-\frac{4E\left(iU_2,2\right)}{\pi^3R_1R_2N}+\frac{32E\left(iU_2,3\right)}{\pi^5R_1^2R_2^2}\right].
		\end{equation}
		
		\subsubsection{Sunset diagram}	
		When the untwisted propagator in the sunset diagram is a scalar we obtain the result
		\begin{align}
		\begin{split}
		I_s&=-\frac{1}{(16\pi^2)^{2}}\sum_{m_i\in\mathbb{Z}}\int_{0}^\infty dy_1dy_2dy_3e^{-y_3m_s^2}\times\frac{2y_3}{(y_1y_2+y_1y_3+y_2y_3)^{3}}e^{-N\left(\frac{1}{y_2}+\frac{1}{y_3}\right)}
		\\
		&=-\frac{1}{(16\pi^2)^{2}}\left(\frac{\pi R_1R_2}{12N^3}-\frac{16}{\pi^5R_1^2R_2^2}\left[\left(3+2\log\frac{N}{\pi^2}\right)E(iU_2,3)+E^{(0,1)}(iU_2,3)\right]\right.
		\\
		&\left.\quad-\frac{4E(iU_2,2)}{\pi^3R_1R_2N}+\frac{32E(iU_2,3)}{\pi^5R_1^2R_2^2}\right),
		\end{split}
		\end{align}
		where the notation $E^{(0,1)}(U,n)\equiv\partial_nE(U,n)$. On the other hand when the untwisted propagator is a fermion we have
		
		\begin{align}
		\begin{split}
		I_f&=-\frac{1}{(16\pi^2)^{2}}\sum_{m_i\in\mathbb{Z}}\int_{0}^\infty dy_1dy_2dy_3e^{-y_2m_f^2}\times\frac{2y_3}{(y_1y_2+y_1y_3+y_2y_3)^{3}}e^{-N\left(\frac{1}{y_2}+\frac{1}{y_3}\right)}
		\\
		&=-\frac{1}{(16\pi^2)^{2}}\left(\frac{\pi R_1R_2}{6N^3}+\frac{16}{\pi^5R_1^2R_2^2}\left[\left(3+2\log\frac{N}{\pi^2}\right)E(iU_2,3)+E^{(0,1)}(iU_2,3)\right]\right).
		\end{split}
		\end{align}
		Therefore the total contribution from the sunset diagrams with unbroken supersymmetry is
		\begin{equation}
		I_s+I_f=-\frac{1}{(16\pi^2)^{2}}\left\{\frac{\pi R_1R_2}{4N^3}-\frac{4E(iU_2,2)}{\pi^3R_1R_2N}+\frac{32E(iU_2,3)}{\pi^5R_1^2R_2^2}\right\}
		\end{equation}
		which exactly cancels the contribution from the figure 8 diagram as expected.
		
		Finally we can obtain the two-loop contribution to the vacuum energy from the twisted diagrams in a theory with supersymmetry broken by the Scherk-Schwarz mechanism. The masses of the twisted states themselves are unaffected by the supersymmetry breaking, but the masses of the untwisted states to which they couple may still be shifted. The result of Scherk-Schwarz supersymmetry breaking amounts to shifting the KK numbers by $\frac{1}{2}$. We may proceed with the calculation in the same way as before, and find the shift in the KK numbers results in a replacement of the real analytic Eisenstein series $E(U,n)$ by $E_{\frac{1}{2}}(U,n)$, where
		\begin{equation}
		E_{\frac{1}{2}}(U,n)=\sideset{}{'}\sum_{\ell_1,\ell_2}\frac{U_2^ne^{\pi i(l_1+l_2)}}{\abs{\ell_1+\ell_2U}^{2n}}\, .
		\end{equation}
		Therefore, we find the contribution from the twisted sectors to be
		\begin{equation}\Tr(Y_{UTT}^2)N^T
		\frac{\left(N_b^U-N_f^U\right)}{16\pi^9R_1^2R_2^2}\left[\left(3+2\log\frac{N}{\pi^2}\right)\tilde{E}(iU_2,3)+\tilde{E}^{(0,1)}(iU_2,3)\right]
		\end{equation} 
		where $\tilde{E}(U,n)$ is an Eisenstein series restricted to $l_1+l_2=$ odd, $N^T$ is the number of twisted degrees of freedom, and $N_b^U$ and $N_f^U$ denote the number of untwisted bosons and fermions respectively that couple to the twisted states and whose masses remain unshifted after supersymmetry breaking. Therefore, we see that if the spectrum contains a degeneracy in the number of massless bosons and fermions in the untwisted sector that couple to twisted states, then the leading contribution from the twisted sectors is zero. Noting that the functional form of this term makes it 
		unnatural for it to cancel against the entirely untwisted contribution, this gives us a second criterion for the vanishing of the two-loop cosmological constant: $\beth=0$ where in terms of the 
		couplings we have 
		\begin{equation}
		\label{eq:master4}
		\beth ~=~ \sum_{U={\rm massless}} (-1)^{F_U} {\rm Tr} |Y_{UTT}|^2 \, ,
		\end{equation}
and where for a given $U$, the coupling $Y_{UTT}$ is considered to be a matrix with indices running over all the twisted states, and includes both gauge and Yukawa couplings. 	
		Taking account of the degeneracy in the couplings, we can write a simple operational expression for $\beth$, namely  
		\begin{equation}
		\beth ~=~ \sum_{U,T,T'={\rm massless}} (-1)^{F_U} \, \delta_Q ({\bf Q}_U+{\bf Q}_T+{\bf Q}_{T'} ) \, ,
		\end{equation}
where the sum is over all massless physical untwisted fields, and pairs of twisted fields. The $\delta_Q$-function imposes {\em either}  simple charge conservation for the 
charge vectors of the triplet of fields (i.e. representing superpotential $\phi \bar{\psi}_L\psi'_R$ type couplings), {\em or} charge conservation with an 
extra unit in the non-compact space-time index (representing gauge $A^\mu \bar{\psi}_L \gamma_\mu \psi'_L$ type couplings that have an extra Dirac matrix).

\section{Conclusion} 

In this paper we have derived two criteria for the exponential suppression of the two-loop cosmological constant in string theories with spontaneously broken 
supersymmetry. These two criteria determine respectively when the leading order entirely untwisted and partially twisted contributions vanish.
The untwisted criterion, in eq.(\ref{eq:master2}), is most easily determined in any given model from the vanishing of the constant term in the $q$-expansion of the 
integrand in the two-loop cosmological constant. Note that this object contains factors from the partition function but also from the measure; the criterion can not  
be determined from the partition function alone. The twisted criterion can be determined from the effective field theory, but can most easily be evaluated in a very simple operational
way simply with the knowledge of the states in the spectrum and all of their charges. The resulting condition, in eq.(\ref{eq:master4}), is the vanishing of a ``sum of Veltman conditions'' for the twisted fields; 
that is, in terms of the effective field theory, one can imagine that at the one-loop level the twisted states in the spectrum will receive quadratically divergent contributions to their mass
from the leading quadratic divergence in the cosmological constant. At the two-loop level, these terms will enter into ``sunset'' diagrams, but the degenerate nature of the 
couplings implies that the sum of such contributions may vanish, depending on the spectrum. 

For consistency, one should also impose the vanishing of the one-loop 
leading contribution to the cosmological constant, which is achieved in theories that have Bose-Fermi degeneracy in their massless physical states. Divergences associated with the one-loop dilaton tadpole would appear at two loop level in the so-called separating degeneration limit of the diagrams, a limit that resembles two one-loop torus diagrams connected by a
 long thin tube.  However, their presence does not actually affect the phenomenology of these models since the crucial point is that because the tadpoles are exponentially suppressed, their effect on the physical spectrum is in fact negligible.
 
 The two criteria we have presented here can be thought of as a stringy implementation of the ``naturalness without supersymmetry'' idea first proposed in ref.\cite{Jack:1989tv} up to the two-loop level. 
The existence or otherwise of models that satisfy these conditions, and their properties should they exist, is a subject of current study, which will be reported elsewhere \cite{future}. 

It would also be of interest to search for a subset of theories that mimic the supertrace rules in models 
involving D3-branes, where vanishing one-loop supertraces are known to extend to higher order automatically \cite{Bena:2015qfa}.

\subsection*{Acknowledgements} We are grateful to Iosif Bena, Keith Dienes, Emilian Dudas, Mariana Gra\~na and Herv\'e Partouche for enlightening discussions, and to Boris Pioline for useful comments. SAA thanks the Ecole Polytechnique for hospitality during the completion of this work. RJS is funded by an EPSRC studentship.

\newpage
\begin{appendices}
	\section{Two-loop theta functions}
	\label{app:a}
	Letting $\tau_{11}\equiv Y_1+Y_2,\ \tau_{12}\equiv Y_2,\ \tau_{22}\equiv Y_2+Y_3$ and defining $q_I=\exp\{2\pi i Y_I\}$, the genus two theta functions have the following expansions in $q_I$ up to linear order (note that the convention for cycles, ${\bf \colt{\alpha_1 V}{\alpha_2 V}{\beta_1 V}{\beta_2 V}}$, is the transpose of that used in \cite{DP1,DP2,DP3,DP4})
	\begin{align}
	\begin{split}
	&\vartheta\colt{0}{0}{0}{0}\sim 1+2q_1^{1/2}q_2^{1/2}+2q_1^{1/2}q_3^{1/2}+2q_2^{1/2}q_3^{1/2}+\ldots \\
	&\vartheta\colt{0}{0}{0}{\frac{1}{2}} \sim 1+2q_1^{1/2}q_2^{1/2}-2q_1^{1/2}q_3^{1/2}-2q_2^{1/2}q_3^{1/2}+\ldots \\	
	&\vartheta\colt{0}{0}{\frac{1}{2}}{0} \sim 1-2q_1^{1/2}q_2^{1/2}-2q_1^{1/2}q_3^{1/2}+2q_2^{1/2}q_3^{1/2}+\ldots\\
	&\vartheta\colt{0}{0}{\frac{1}{2}}{\frac{1}{2} } \sim 1-2q_1^{1/2}q_2^{1/2}+2q_1^{1/2}q_3^{1/2}-2q_2^{1/2}q_3^{1/2}+\ldots \\
	&\vartheta\colt{\frac{1}{2}}{0}{0}{0}\sim 2q_1^{1/8}q_2^{1/8}(1+q_3^{1/2})+\ldots\\
	&\vartheta\colt{\frac{1}{2}}{0}{0}{\frac{1}{2}} \sim 2q_1^{1/8}q_2^{1/8}(1-q_3^{1/2})+\ldots\\
	&\vartheta\colt{0}{\frac{1}{2}}{0}{0} \sim 2q_2^{1/8}q_3^{1/8}(1+q_1^{1/2})+\ldots\\
	&\vartheta\colt{0}{\frac{1}{2}}{\frac{1}{2}}{0}\sim 2q_2^{1/8}q_3^{1/8}(1-q_1^{1/2})+\ldots\\
	&\vartheta\colt{\frac{1}{2}}{\frac{1}{2}}{0}{0}\sim 2q_1^{1/8}q_3^{1/8}(1+q_2^{1/2})+\ldots\\
	&\vartheta\colt{\frac{1}{2}}{\frac{1}{2}}{\frac{1}{2}}{\frac{1}{2}}\sim 2q_1^{1/8}q_3^{1/8}(1-q_2^{1/2})+\ldots
	\end{split}
	\end{align}
	For ease of reference we also collect here the large radius $q$-expansion for the weight 10 Igusa cusp form:
	\begin{equation}
	\frac{2^{12}}{\Psi_{10}}=\frac{1}{q_1q_2q_3}+2\sum_{i<j}\frac{1}{q_iq_j}+24\sum_i\frac{1}{q_i}+\mathcal{O}(q_i)\, .
	\end{equation}
		
	\section{Modular transformations for genus 2 surfaces}
	Modular transformations for a genus 2 Riemann surface form the infinite discrete group $\Sp(4,\mathbb{Z})$ defined by
	\begin{equation}
	M=
	\begin{pmatrix}
	A & B
	\\
	C & D
	\end{pmatrix},
	\qquad M
	\begin{pmatrix}
	0 & I
	\\
	-I & 0
	\end{pmatrix}
	M^t=
	\begin{pmatrix}
	0 & I
	\\
	-I & 0
	\end{pmatrix}\, ,
	\end{equation}
	where $A,B,C,D$ are integer valued $2\times 2$ matrices. The Siegel upper half-plane is defined as the set of all symmetric $2\times2$ complex matrices with positive definite imaginary part. Modular transformations under $\Sp(4,\mathbb{Z})$ act on the Siegel upper half-plane by
	\begin{equation}
	\Omega\rightarrow\tilde{\Omega}=(A\Omega+B)(C\Omega+D)^{-1},
	\end{equation}
	giving the following transformations,
	\begin{align}\label{ModT}
	\begin{split}
	\vartheta[\tilde{\delta}](0,\tilde{\Omega})^4&=\epsilon^4\det(C\Omega+D)^2\vartheta[\delta](0,\Omega)^4,
	\\
	\Xi_6[\tilde{\delta}](\tilde{\Omega})&=\epsilon^4\det(C\Omega+D)^6\Xi_6[\delta](\Omega),
	\\
	\Psi_{8}(\tilde{\Omega})&=\det(C\Omega+D)^{8}\Psi_{8}(\Omega),
	\\
	\Psi_{10}(\tilde{\Omega})&=\det(C\Omega+D)^{10}\Psi_{10}(\Omega),
	\\
	\det\Im(\tilde{\Omega})&=\abs{\det(C\Omega+D)}^{-2}\det\Im\Omega,
	\\
	d^3\tilde{\Omega}&=\abs{\det(C\Omega+D)}^{-6}d^3\Omega,
	\end{split}
	\end{align}
	where $\epsilon^4=\pm1$.
	
	\section{$SO(10)$ model with massless Bose-Fermi degeneracy}
	\label{so10}
       \subsection{Model definition, and vanishing of SUSY partition function}
		The model is defined by the following set of basis vectors ${\bf V}_a$ and CDC deformation vector ${\mathbf e}$, which correspond to the $SO(10)$ model of ref.\cite{ADM}:
		\begin{align}
		\begin{split}
		\bm{V}_0&=-\frac{1}{2}[11\ 111\ 111\ |\ 1111\ 11111\ 111\ 11111111]
		\\
		\bm{V}_1&=-\frac{1}{2}[00\ 011\ 011\ |\ 1111\ 11111\ 111\ 11111111]
		\\
		\bm{V}_2&=-\frac{1}{2}[00\ 101\ 101\ |\ 0101\ 00000\ 011\ 11111111]
		\\
		\bm{b}_3&=-\frac{1}{2}[10\ \bar{1}0\bar{0}\ \bar{0}0\bar{1}\ |\ 0001\ 11111\ 010\ 10011100]
		\\
		\bm{V}_4&=-\frac{1}{2}[00\ 101\ 101\ |\ 0101\ 00000\ 011\ 00000000]
		\\
		\bm{e}&=\phantom{-}\frac{1}{2}[00\ 101\ 101\ |\ 1011\ 00000\ 000\ 00011111]\, ,
		\end{split}
		\end{align}
		while the corresponding structure constants $k_{ij}$ are given by
		\begin{equation}
		k_{ij}=\begin{pmatrix}
		0 & 0 & 0 & \frac{1}{2} & 0
		\\
		0 & 0 & 0 & \frac{1}{2} & 0
		\\
		0 & \frac{1}{2} & 0 & 0 & 0
		\\
		\frac{1}{2} & 0 & 0 & 0 & 0
		\\
		0 & \frac{1}{2} & 0 & 0 & 0
		\end{pmatrix}\, .
		\end{equation}
		It is easier to verify the vanishing of the two loop cosmological constant in SUSY models  by taking a set of equivalent basis vectors where $V_0$ and $V_1$ are replaced by
		\begin{align}
		\begin{split}
		\bm{V}_0'=\bm{V}_1&=-\frac{1}{2}[00\ 011\ 011\ |\ 1111\ 11111\ 111\ 11111111]
		\\
		\bm{V}_1'=\overline{\bm{V}_0+\bm{V}_1}&=-\frac{1}{2}[11\ 100\ 100\ |\ 0000\ 00000\ 000\ 00000000]\, .
		\end{split}
		\end{align}
		Beginning with a simple model defined only by the vectors ${\bf V}_0'$ and ${\bf V}_1'$, one finds a contribution appearing as an overall factor in the expression for the cosmological constant. This factor comes from the components corresponding to $i_R=1,2,3,6$ and is given by
	\begin{equation}
	\sum_{a,b,c,d\in\{0,\frac{1}{2}\}}\Xi_6\colt{a}{b}{c}{d}\vartheta\colt{a}{b}{c}{d}^4=0\, .
	\end{equation}
	A similar story applies to the model defined by the three basis vectors ${\bf V}_0'$, ${\bf V}_1'$ and ${\bf V}_2$ where the identity that now guarantees a vanishing cosmological constant is
	\begin{equation}
	\sum_{a_1,b_1,c_1,d_1\in\{0,\frac{1}{2}\}}(-1)^{c_2a_1+d_2b_1}\Xi_6\colt{a_1}{b_1}{c_1}{d_1}\vartheta\colt{a_1}{b_1}{c_1}{d_1}^2\vartheta\colt{a_1+a_2}{b_1+b_2}{c_1+c_2}{d_1+d_2}^2=0\, ,
	\end{equation}
	for any $a_2,b_2,c_2,d_2\in\{0,\frac{1}{2}\}$.
	By inspection, this identity also guarantees a vanishing contribution to the one-loop vacuum energy of the full non-SUSY $SO(10)$ model above, from the untwisted sectors in which both $\ell^1,\ell^2=0\mod{(2)}$ (where $\ell^1=\ell^1_1+\ell^1_2$ and similar for $\ell^2$). 
	
\subsection{Massless Bose-Fermi degeneracy and the 1-loop $q$-expansion}
	The one-loop partition function after the applying the CDC is proportional to
	\begin{equation}
	\mathcal{Z}\propto\frac{1}{\eta(\tau)^{12}\bar{\eta}(\bar{\tau})^{24}}\sum_{\alpha,\beta}C\col{\alpha}{\beta}\Gamma_{2,2}\bigg|_{n=0}\prod_{i_R}\vartheta\col{\overline{\alpha \bm{V}_i}-n\bm{e}_i}{-\beta\bm{V}_i+\ell\bm{e}_i}\prod_{i_L}\bar{\vartheta}\col{\overline{\alpha \bm{V}_i}-n\bm{e}_i}{-\beta\bm{V}_i+\ell \bm{e}_i}\, .
	\end{equation}
	The $q$-expansions of $\eta(\tau)^{-12}$ and $\bar{\eta}(\bar{\tau})^{-24}$ are
	\begin{equation}
	\begin{split}
	\frac{1}{\eta(\tau)^{12}}&=\frac{1}{\sqrt{q}}+\mathcal{O}(\sqrt{q}),
	\\
	\frac{1}{\bar{\eta}(\bar{\tau})^{24}}&=\frac{1}{\bar{q}}+24+\mathcal{O}(\bar{q})\, .
	\end{split}
	\end{equation}
	The source of the exponential suppression of the one loop cosmological constant is then that, in the sectors where  $\ell=\ell_1+\ell_2$ is odd (so that the contributions does not just vanish by supersymmetry), the $q$-expansion of the partition function is found to be missing the constant term due to the Bose-Fermi degeneracy among the massless states:
	\begin{align}
	\begin{split}
	\mathcal{Z}&\propto\frac{1}{\eta(\tau)^{12}\bar{\eta}(\bar{\tau})^{24}}\left(128\sqrt{q}-3072\bar{q}\sqrt{q}+\ldots\right)
	\\
	&=\frac{128}{\bar{q}}+0+\mathcal{O}((q\bar{q})^{1/2})\, .
	\end{split}
	\end{align}

	\end{appendices}

\newpage


\end{document}